\documentclass[a4paper,11pt]{article}
\usepackage{jinstpub} 
\usepackage{lineno}

\title{\boldmath AI for Nuclear Physics: the EXCLAIM project}

\author[a]{S. Liuti, \note{Corresponding author}}
\author[a]{D. Adams}
\author[b]{M. Bo\"{e}r}
\author[a]{G.W. Chern}
\author[a]{M. Cuic}
\author[c]{M. Engelhardt}
\author[d]{G.R. Goldstein}
\author[e, *]{B. Kriesten}
\author[f]{Y. Li}
\author[g]{H.W. Lin}
\author[c]{M. Sievert}
\author[h, *]{D. Sivers}

\affiliation{a Physics Department, University of Virginia,
382 McCormick Rd, Charlottesville, VA 22901, USA}
\affiliation{b Department of Physics, Virginia Tech,
Blacksburg, VA 24061, USA}
\affiliation{c
Department of Physics, New Mexico State University (NMSU)
Box 30001 - MSC 3D Las Cruces, NM 88003, USA}
\affiliation{d Department of Physics and Astronomy, Tufts University, 
Robinson Hall, Medford, MA 02155, USA}
\affiliation{e Argonne National Laboratory,
9700 S. Cass Avenue, Argonne, IL 60439}
\affiliation{f Department of Computer Science
Old Dominion University Norfolk, VA 23529, USA}
\affiliation{g Department of Physics and Astronomy, Michigan State University
3218 Biomedical Physical Science Building, East Lansing, MI 48824, USA}
\affiliation{h Department of Physics, University of Michigan
450 Church St., 2477 Randall Lab, Ann Arbor, MI 48109-1040, USA}
\affiliation{* Affiliate}

\emailAdd{sl4y@virginia.edu}

\abstract{An overview of the recent activity of the newly funded EXCLusives with AI and Machine learning (EXCLAIM) collaboration is presented. 
The main goal of the collaboration is to develop a framework to implement AI and machine learning techniques in problems emerging from the phenomenology of high energy exclusive scattering processes from nucleons and nuclei, maximizing the information that can be extracted from various sets of experimental data, while implementing theoretical constraints from lattice QCD. 
A specific perspective embraced by EXCLAIM is to use the methods of theoretical physics to understand the working of ML, beyond its standardized applications to physics analyses which most often rely on industrially provided tools, in an automated way. }


\begin{document}
\maketitle
\flushbottom

\section{Introduction}
\label{sec:intro}
A new generation of experiments at Jefferson Lab and at the Electron Ion Collider (EIC) \cite{AbdulKhalek:2021gbh} -- in conjunction with 
other laboratories worldwide -- will measure and further explore the internal structure of the proton, neutron and atomic nuclei. The most ambitious goal of these efforts will be to provide a mapping of the 3D spatial structure of all strongly interacting systems at the subatomic level along with a deeper understanding of the origin of their mass and spin. 
The nuclear theory community is correspondingly poised to develop new approaches including, most prominently the computational sector, that will allow us to capitalize on the data emerging from these experiments. To this end, we formed a new, multi-disciplinary research group , the ``EXCLusives with Artificial Intelligence and Machine learning" (EXCLAIM) collaboration. 

The major goals of EXCLAIM is to devise a data analysis that combines the experimental information with first-principles theoretical calculations, through the use of modern machine learning (ML) techniques. We believe that such a concerted effort involving experts from computer science, phenomenology, nuclear theory, lattice QCD and experiment, is necessary in order to enable a quantitative and physically sound 3D picture of the nucleon, and to explore its phenomenological consequences. 
The new methods that we are introducing are centered on designing physics informed deep learning architectures where physics constraints are inherently satisfied in their predictions. Both theoretical and experimental constraints
are accounted for in an advanced  statistical framework, as stringently and consistently as possible, by providing the technology to continually incorporate both the latest experimental data and precision
first-principle lattice QCD calculations. 
Unlike standard ``black box'' methods, the new algorithms are crafted for specific physics purposes using more general ML building blocks. 
By generating these algorithms within our joint effort,  working together in an interdisciplinary team, the EXCLAIM group aims at creating a state of the art computational framework that is configured for the specific purpose of learning hadronic structure from experimental data.
%
\section{The importance of  deeply virtual exclusive processes}
Deeply virtual exclusive processes, $ep \rightarrow e'p' \gamma (M) $,
where either a photon or a meson are produced off a proton target with a large four-momentum transfer square, $Q^2$, between the initial and final leptons, were proposed to measure generalized parton distributions (GPDs) \cite{Ji:1996ek,Ji:1996nm} (for reviews see Refs.\cite{Diehl:2003ny,Belitsky:2005qn,Kumericki:2016ehc}). GPDs allow us to go beyond the standard partonic longitudinal momentum-based description of nucleon dynamics, accessed through inclusive deep inelastic scattering, and explore spatial degrees of freedom through Fourier transformation in the momentum transfer between the initial and final proton. Orbital angular momentum which requires the knowledge of the struck quark/gluon spatial coordinate, can also be accessed through GPDs: its measurement would contribute fundamentally to our understanding of the nucleon spin sum rule.   

\begin{figure*}
    \centering
    \includegraphics[width=7.5cm]{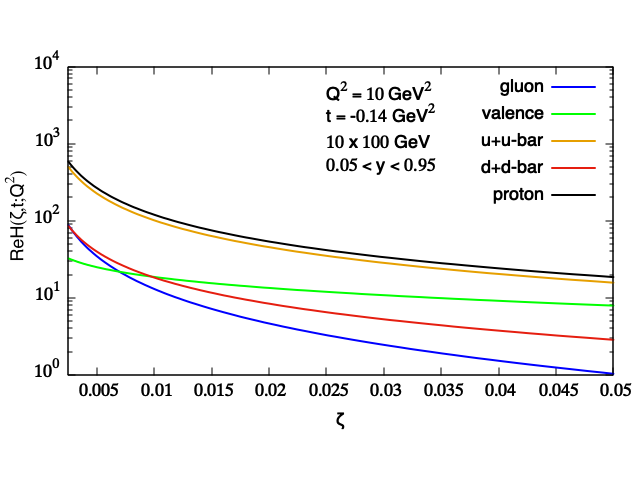}
   \includegraphics[width=7.5cm]{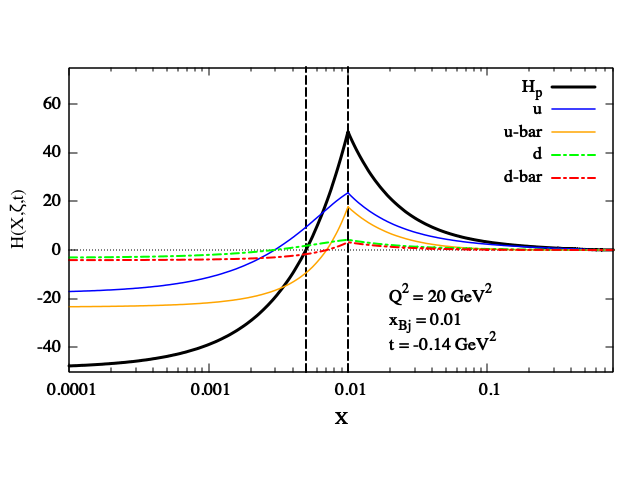}
    \caption{Contributions to the proton CFF as a function of $x_{Bj} = \zeta$ at a fixed kinematics $t = -0.14$ GeV$^{2}$. (\textit{Left}) The proton CFF is separated into gluon, valence, $u+\bar{u}$, and $d+\bar{d}$ components, where all components are scaled by their charges $e_{u}^{2} = 4/9$ and $e_{d}^{2} = 1/9$ at a fixed $Q^{2} = 10$ GeV$^{2}$. (\textit{Right}) GPD $H$ contributing to the CFF (notice that $Q^2 = 20$ GeV$^2$, value is slightly higher than the one used to calculate the CFF. This is however, inconsequential for our description.)}
    \label{fig:cff_vxbj_eic}
\end{figure*}

Producing a photon in the final state as in Deeply virtual Compton scattering (DVCS), provides a cleaner probe than in meson production (DVMP),  because no additional hadronic composite particle is present in the final state. However,
DVCS is affected by a large background, the Bethe-Heitler (BH) process, where the final photon is radiated from the electron, and hadronic structure is described by the Dirac and Pauli form factors as in elastic electron-proton scattering.  
Obtaining GPDs from data presents several other challenges. The perturbative QCD structure of the observables entering the amplitudes for the DVCS and DVMP processes is given by convolutions of the GPDs with complex QCD Wilson coefficient functions over the longitudinal momentum fraction $x$, known as Compton Form Factors (CFFs). The extraction of the full $x$ dependence of GPDs from CFFs requires defining strategies to overcome an inverse problem. 
Furthermore, even at leading order in the QCD twist expansion, all CFFs from the various quark-proton polarization configurations allowed by parity and charge conjugation invariance contribute to the cross section coherently, {\it i.e.} they are summed at the amplitude level. This situation mandates a meticulous analysis for the extraction from data, whose benchmarking has still not been finalized by the community. The helicity amplitudes formalism, illustrated in Refs.\cite{Kriesten:2019jep,Kriesten:2020apm}, clearly highlights this problem. 
In Figure \ref{fig:cff_vxbj_eic}(left) we show the contributions of the gluon, valence, $u+\bar{u}$, and $d+\bar{d}$ components to the CFF for the proton GPD $H$, plotted vs. the skewness variable, $\zeta \approx x_{Bj}$, for a typical EIC kinematic setting, with electron beam energy, $E_e= 10$ GeV and proton beam energy, $E_p= 100$ GeV, momentum transfer $t=-0.2$ GeV$^2$ and scale, $Q^2= 10$ GeV$^2$; in Figure \ref{fig:cff_vxbj_eic}(right) we show the proton GPD $H$ (black line) contributing to the CFF. One can see that the process is dominated by the $\bar{u}$ quark contribution (yellow line on left, and blue line on right). The dashed vertical lines in the figure demarcate the ERBL region, for $X \leq\zeta=0.01$, and the point, $X= \zeta/2$, around which the valence, $q^-=q-\bar{q}$, and $q^+=q+\bar{q}$ distributions display a symmetric behavior in $X$.   

\section{Framework for a Global Analysis of Deeply Virtual Exclusive Processes}
\begin{figure}
    \centering
    \includegraphics[width=6.5cm]{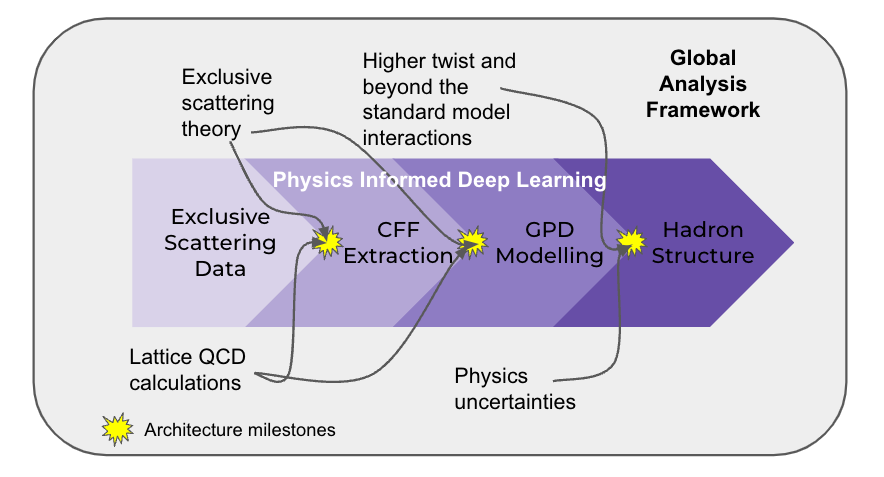}
    \caption{Pipeline of physics informed deep learning framework that goes from exclusive scattering data to information on hadron structure.}
    \label{fig:pipeline}
\end{figure}

In order to obtain the GPDs as well as any other useful physical information on the proton 3D structure from data, quantitative analyses with a faithful description of uncertainty need to be conducted, starting from a well defined coherent set of benchmarks. 
A clean, statistically significant extraction of 3D spatial quark and gluon distributions from data requires, however, a change in paradigm \cite{Almaeen:2022imx,Grigsby:2020auv} from the traditional analytic and computational methods used thus far \cite{Kumericki:2007sa,Moutarde:2009fg,Guidal:2009aa,Cuic:2020iwt}, which after much published work spanning two decades, are still inconclusive.
%
In our analysis we build on the framework for benchmarking the DVCS cross section and, more generally the phenomenology of deeply virtual exclusive scattering, proposed in \cite{Almaeen:2022imx} to address the precision extraction of GPDs in a range of experimental settings, going from Jefferson Lab@6 GeV and 12 GeV, to the EIC 
for typical settings currently projected at the highest  luminosity \cite{AbdulKhalek:2021gbh}. 
Using the helicity amplitudes-based formalism introduced in \cite{Kriesten:2019jep,Kriesten:2020apm,Kriesten:2020wcx} we adopt a streamlined format of the DVCS-BH cross section where the phase given by the azimuthal angle, $\phi$, between the plane containing the initial and final electrons (lepton plane) and the plane containing the final hadron and photon is used to disentangle the QCD dynamical (higher twist) and kinematical power correction terms in the scale, $1/Q$. A thorough understanding of the scale dependence of the cross section is key for understanding what features of GPDs, {\it e.g.} their flavor composition, symmetries, low $x$ vs. large $x$ behavior, can be determined in the present and future experimental settings.
%
%
The contribution of twist-three GPDs needs to be carefully evaluated, 
not only to gauge their contamination of the twist-two terms at energy scales of a few GeV, but also because they provide  essential information on the emergence of proton Orbital Angular Momentum (OAM) \cite{Polyakov:2002yz,Polyakov:2018zvc} and mass \cite{Ji:2021qgo,Ji:1995sv,Lorce:2018egm} through color forces \cite{Burkardt:2012sd}, specifically quark-gluon-quark interaction terms in the off-forward correlation function \cite{Raja:2017xlo,Rajan:2016tlg}. 
%
%
A schematic representation of our approach with a physics informed deep learning framework that goes from exclusive scattering data to information on hadron structure is presented in Figure \ref{fig:pipeline}.
%
%

%
%

\subsection{Extraction of Compton form factors from experiment}
Numerous steps are involved to obtain CFFs starting from raw experimental data. CFFs are functions of GPDs that can be accessed from experimental observables. 
Experimental data come in large data sets that need, after various other steps of the data analysis, to be singled out from the physics background given by other exclusive and semi-inclusive channels.
Proper background reduction and control of the systematic uncertainties necessitates efficient analysis algorithms, and the use of models and simulations for all of the potentially contaminating channels. It is a well known fact that some of the background cannot be reduced, and one has to  estimate its contribution to the observable of interest  (cross section or asymmetry). There are growing efforts towards implementing and using ML techniques in these specific aspects of the data analyses, contributing to a better background reduction  and to the speed of the algorithms. Our research is poised to improve the  understanding and control of systematic uncertainties introduced by these methods. Through a close coordination between experimentalists and theorists we use accurate models including parameters for the main and background channels to reproduce experimental results.  Another essential focus of our analysis through a coordinated experimental/theory effort, is to determine which observable are more likely to be measured accurately. The subsequent step of our analysis, after the CFFs are extracted from the various cross sections, asymmetries and moments of asymmetries, concerns the error determination. Depending on how the data are presented, part of the information needed to  propagate the uncertainties to the extracted CFFs can be lost. In our approach, we optimize the extraction of CFFs and GPDs accounting for the background, within the phase space and acceptance limitations of the experiments. 
As we explain below, we also explore the use of the complementary information given by lattice QCD predictions, to maximize the extraction of information on 3D structure from data.
Algorithms that weigh in nonlinear correlations among data such as in a VAIM framework have already been used to understand similar inverse problems such as finding the parameters in PDF fits of DIS data. We are further developing unsupervised algorithms to make use of the physics information carried in the experimental errors.  

\subsection{AI-assisted global fitting of experimental and lattice-QCD data.} 
\begin{figure}[http]
\begin{center}
      \includegraphics[scale = 0.30]{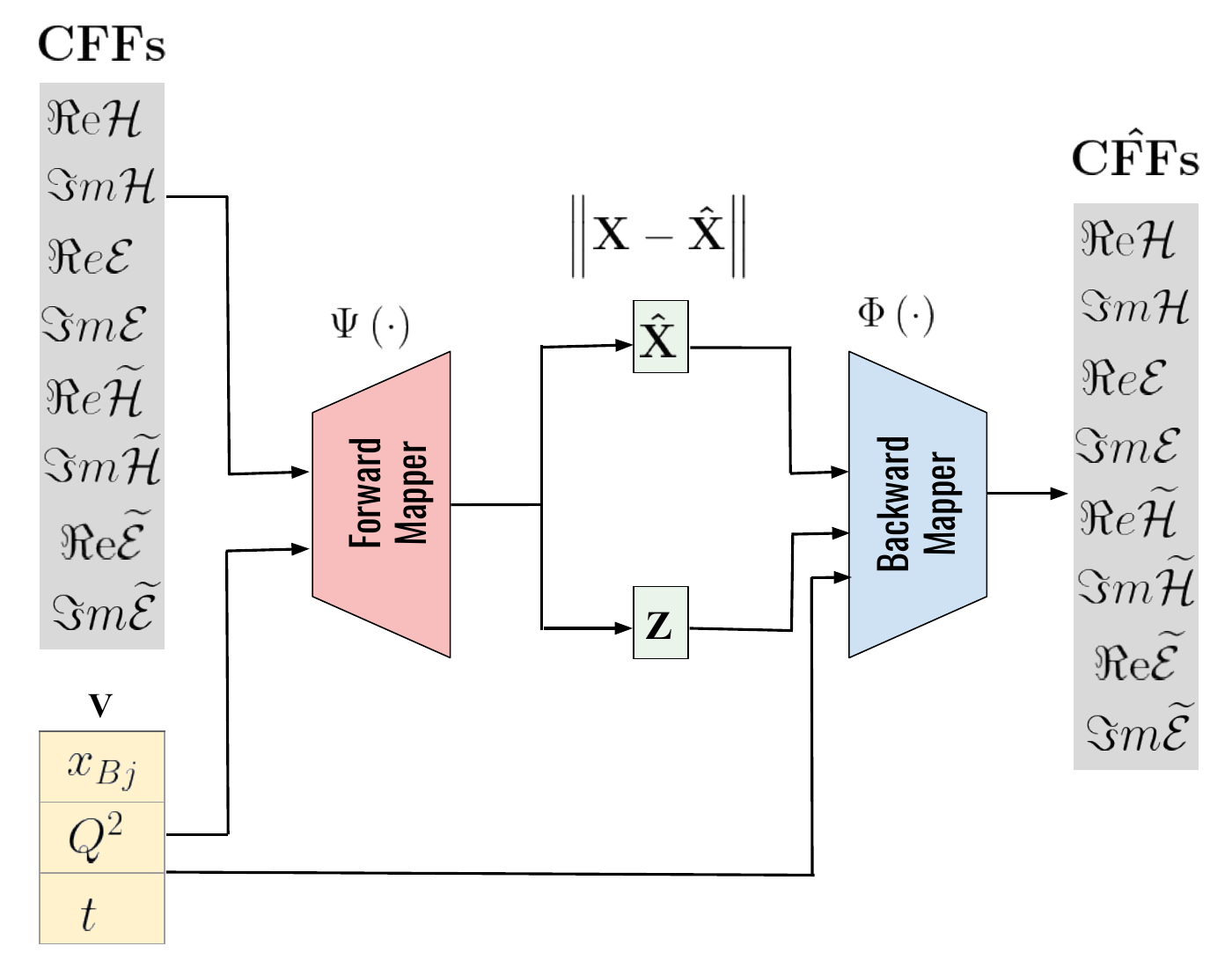}
    \caption{Overview of C-VAIM architecture}
    \label{fig:vaim_arch}
    \end{center}
\end{figure}

Our collaboration is developing an open-source, comprehensive global analysis phenomenology package for 3D imaging and dynamical hadron tomography, which will provide accessible and flexible tools to analyze and interpret the soon to be available EIC data  \cite{Khan:2022vot}. Our approach is based on a rigorous theoretical and statistical framework making use of all constraints from QCD, including Lattice QCD. 
The EXCLAIM collaboration leverages physics informed neural networks to solve inverse problems in extracting information on hadronic structure from experimental data. These deep learning (DL) models can be implemented in two ways:  1) by imposing soft constraints in which the DL model is ``suggested" to satisfy certain physics constraints; and 2) by imposing hard constraints in which the DL model will automatically satisfy constraints in its predictions. Soft constraints are implemented by adding additional terms to the loss function, meaning that the DNN will trend towards minimizing these additional terms but not exactly satisfy them. These added constraint terms are implemented similar to Lagrange multipliers. The flexibility of soft constraints can be useful when working with real experimental data which includes stochastic noise. Hard constraints are implemented into the architecture of the DNN itself such as convenient choices of model architecture size, invertibility of the model architecture, and clever choice of activation functions. 
The DL methods we have been implementing have been so far based on a Variational Autoencoder Inverse Mapper (VAIM)~\cite{Almaeen:2021ijcnn} 
to solve the inverse problem in CFF extraction from a single polarization observable. The VAIM is an end-to-end deep learning framework (as shown in Figure \ref{fig:vaim_arch}), employing an autoencoder-based deep neural network architecture to address the ambiguity issue in inverse problems. The encoder and decoder neural networks approximate the forward and backward mapping, respectively, and a variational latent layer is incorporated into VAIM to learn the posterior parameter distributions with respect to given observables. The fundamental idea of VAIM is to use the variables in the variational latent layer to capture the lost information in forward mapping and convert an ill-posed inverse problem into a well-posed regression problem. When the experimental data are observed on irregular and varying kinematics bins, VAIM can be extended to a point cloud-based VAIM (PC-VAIM), with a permutation invariant neural network framework to handle the observables in unstructured and unordered point cloud representations. 
In our previous work, VAIM has been successful in determining multiple solutions to the parameterization of quark/anti-quark/gluon parton distribution functions from structure function measurements. Through the use of the VAIM framework, one can reconstruct the lost information in forward mapping through analysis of the latent space variables, which converts the ill-posed inverse problem into a well-posed regression problem. Given the cross section as input, sampling the latent layer can reconstruct the lost information and consequently determine the corresponding parameter distribution of the parton distribution function.  
We targeted a similar goal in the context of DVCS. Treating the experimental observable as an equation that is parametrized by CFFs, we can recast the VAIM framework to study the effect of extracting CFFs with potentially multiple solutions. 
Through the application of regulated data augmentation methods, we will be able to then propagate experimental error into the VAIM model to quantify the uncertainty associated with CFF parameterization, which may prove beneficial to analyzing exclusive scattering experiments.

The following physics constraints are introduced into our VAIM framework at various stages:
1) {Positivity} constraints from GPDs \cite{Martin:1997wy,Pire:1998nw,Diehl:2000xz}; 
2) {Dispersion relations}between $\Re e$ and $\Im m$ parts of the extracted CFFs \cite{Anikin:2007yh,Diehl:2007jb}  with proper consideration of threshold effects \cite{Goldstein:2009ks}; 
3) {Rosenbluth separations}: exploiting linear relations in the cross section through transverse and longitudinal separations; 
4) {Higher order corrections} including higher twist contributions including spin-orbit and diquark correlations, as well as $\alpha_{s}$ contributions to the cross section, built into the architecture of the network. 
%
5) Lorentz invariance through polynomiality using elastic scattering data and lattice QCD calculated moments;
6) Symmetries among kinematic variables \cite{Golec-Biernat:1998zbo,Diehl:2003ny};
7)  Forward limit of GPD as defined by $\xi,t \rightarrow 0$ compared to parton distribution function.

The extraction of the full $x$-dependence of the GPD is being attacked by using a multi-stage approach where, in the first stage of the analysis, we use the VAIM framework developed for the extraction of CFFs and modify this architecture into VAIM-GPD. The VAIM-GPD framework takes an input parameterization developed using a helicity amplitude-based spectator model calculation to determine all possible configurations of the parameter space. This first step is necessary to test ensembling all hadronic information from various sources together to best inform a single prediction of the GPD. The use of a parameterization allows us to determine with a best-case scenario if this problem can be solved and will reveal a plethora of information regarding hadronic properties. Included in this analysis is an exploration of various uncertainty quantification techniques.
The latter derive from two main sources in our project: from the data (aleatory), and from the model (epistemic). The first describes the accuracy of the data and it is dominated by a statistical component, the second stems from the model's features and flexibility and it depends on the framework used, 
with models requiring weak dependence between data points. 
It is a measure of how uncertainty, or the variance, can be increased or decreased by the presence of this dependence. 
%
Further studies will allow us to address their separate contributions in the context of a global analysis of deeply virtual exclusive experiments. Similar studies have been performed in the context of global analyses of inclusive scattering experiments ~\cite{Kovarik:2019xvh}.

\section{Conclusions and Outlook}
Recent efforts in both the nuclear physics and computer science communities have been developing AI/ML tools and methods to address the challenges of the increasingly complex outcomes of nuclear physics experiments and to enable unbiased theoretical interpretation. The EXCLAIM collaboration aims at providing an original pathway for the use of such methods in the theoretical analysis of the experimental data on DVES.
by establishing, on one side, a common set of benchmarks for both the ML and physics practitioners, and, on the other, by developing algorithms for the computational analysis of the data.  
What sets our approach apart from a standard neural network analysis is the physics content that is injected in strategic, informed ways at various stages of the analysis. 
Our results so far include using theory constraints, lattice QCD data, and experimental measurements 
as an integral part of the VAIM architecture to extract CFFs and GPDs from available data.
A lack of direct GPD measurements makes a global extraction of GPDs from experimental data a hard problem to solve using standard supervised learning techniques; therefore, an exploration of modern algorithms that blend supervised and unsupervised learning are necessary. 
We will therefore develop, in future efforts,  approach based on reinforcement learning (RL) sequential decision making processes and generative models that are independent of particular parametric form. 
This will lead into a first application of sequence-to-sequence RL methods originating from Natural Language Processing (NLP) \cite{keneshloo2019deep,du2019empirical} 
to learn information on hadronic structure from GPDs that satisfy various experimental and theoretical constraints. 
A successful application of these methods to theoretical nuclear physics provides valuable contributions to both the physics and AI communities. 

\acknowledgments
This work was completed by the EXCLAIM collaboration under the DOE grant DE-SC0024644. 


\bibliographystyle{JHEP}
\bibliography{biblio.bib}


\end{document}